\documentclass[aps,prl,amsmath,twocolumn,showpacs]{revtex4}
\usepackage{graphicx}
\begin{document}
\title{Magnetoelectric effect in mixed valency oxides mediated by charge carriers}
\author{V.A. Stephanovich}\homepage{http://cs.uni.opole.pl/~stef}
\email{stef@math.uni.opole.pl}
\affiliation{Opole University, Institute of Mathematics and
Informatics, Opole, 45-052, Poland}
\author{M.D.Glinchuk}
\email{glin@materials.kiev.ua} \affiliation{Institute for Problems
of Materials Science, National Academy of Science of Ukraine,\\
Krjijanovskogo 3, 03142 Kiev, Ukraine}
\author{R.Blinc}
\email{Robert.Blinc@ijs.si} \affiliation{Josef Stefan Institute,
Jamova 39, 1000 Ljubliana, Slovenia}

\date{\today}
\begin{abstract}
We show that the presence of free carriers in a substance can
generate the multiferroic behavior. Namely, if the substance has
mixed-valence ions, which can supply free carriers and have electric
dipole and spin moments, all three types of long-range order
(ferromagnetic, ferroelectric and magnetoelectric (ME)) can occur at
low temperature. The physical origin of the effect is that charge
carriers can mediate the multiferroic behavior via spin - spin
(RKKY), dipole-dipole and dipole - spin interactions. Our estimate
of the interaction magnitude shows that there exist an optimal
carrier concentration, at which the strength of ME interaction is
maximal and comparable to that of spin-spin RKKY interaction. This
permits to conclude that in substances, where RKKY interaction
between local spins is not small, a substantial value of free
carriers mediated ME interaction can occur. Our analysis shows that
disorder in the above substances does not suppress multiferroic
effects.
\end{abstract}
 \maketitle
Multiferroics are substances which can simultaneously exhibit
spontaneous polarization, magnetization and/or strain \cite{fib}.
The ferroics where magnetic and electric orders coexist have
recently become the focus of intensive research
\cite{nat04,prl1,prl2}, magnetism and ferroelectricity being
involved with local spins and off-center structural distortions
(electric dipoles) respectively. The physical nature of the
coexistence of these two seemingly unrelated phenomena as well as of
the recently observed \cite{prl1} strong interaction between
polarization and magnetization (magnetoelectric (ME) coupling) is
not completely understood up to now, because trivial spin-orbit
interaction can be the source of a very small ME effect only. The
clarification of the physical nature of above phenomena would be of
great importance both for fundamental research and for creation of
new generations of electronic devices. In particular, the ME effect
in multiferroics opens the way to write an information with the help
of the electric field and to read it with the help of the magnetic
field making the memory devices more stable and reliable
\cite{scott}. Among different multiferroics with high ME effect the
oxides seem to be the largest group. Namely, perovskite rare-earth
manganites RMnO$_3$ (R = Eu, Gd, Tb and Dy) \cite{prl1, kimura},
perovskite-type BiFeO$_3$, BiMnO$_3$ (see Ref. \cite{ramesh} and
references therein), materials with general formulae RMn$_2$O$_5$ (R
= Tb, Y, Eu) \cite{nat04,prl2} and other oxides belong to this
group\cite{fib}. It is easy to see that one of these oxides cations
belongs to the ions of 3d-group known to have several valency
states. Such ions are Mn$^{2+}$, Mn$^{3+}$, Mn$^{4+}$, Fe$^{2+}$,
Fe$^{3+}$, Cr$^{3+}$, Cr$^{5+}$ etc. The presence of these ions have
been observed by ESR and optical methods \cite{slikter_book} so that
mixed valency is expected to occur in the above materials. Moreover,
due to chemical and structural complexity of RMn$_2$O$_5$, Mn$^{4+}$
ions are octahedrally coordinated by oxygen, whereas Mn$^{3+}$ are
at the base center of a square pyramid \cite{nat04} so that the
general formula of this oxide can be rewritten as
R(Mn$^{4+}$Mn$^{3+}$)O$_5$ with equal number of Mn$^{4+}$ and
Mn$^{3+}$. So, the cation valency depends on the local symmetry,
defined by the number of surrounding oxygen ions. However it is
known that the oxides used to have a lot of oxygen vacancies (and
possibly other imperfections) randomly distributed in the host
lattice. This can lead to a random distribution of cations with
different valency, some of them can be magnetic or nonmagnetic (if
their ground state spin is zero). On the other hand, oxygen
vacancies can generate free carriers so that the aforementioned
oxides can acquire nonzero conductivity equal to that of
semiconductors. We emphasize here, that this finite conductivity
does not mean complete destruction of possible ferroelectric order,
rather, many interesting effects appear, see \cite{frid} for
details. These effects, however, are not directly related to the ME
effect mechanism, discussed here.

As we have mentioned above, the cation valency determines its
position either in the center of high symmetry surrounding or in the
off-center position. This is because the position is defined by
delicate balance of repulsive and attractive interactions between
the ion and its surroundings, so that smaller ions would be expected
to have the energy minimum in the off-center position (see Ref.
\cite{grimes} and references therein).

In Ref. \cite{grimes}, the data related to off-center positions of
cations in spinel structure were reported. Note, that off-centrality
of an ion can be considered as a result of a pseudo-Jan Teller
effect \cite{glin_karm73}.

It had been shown \cite{dietl97} that in diluted magnetic
semiconductors (DMS) (for instance Ga$_{1-x}$Mn$_x$As), also
belonging to above class of mixed valency substances, ferromagnetism
appears due to Ruderman-Kittel-Kasuya-Yosida (RKKY) indirect
coupling between Mn d-shell localized magnetic moments mediated by
the induced spin-polarization in a free-hole itinerant carrier
system. The experimental investigation of (Ga,Mn)As DNS
\cite{ohno98} shows that the temperature of ferromagnetic phase
transition is $T_c=$110 K or even $T_c=$173 K \cite{jun05}, which is
quite a large value allowing for the randomness of the Mn ions
distribution. Since it is not excluded, that the ion (for example
off-center Mn) can have both spin and electric dipole moment, the
above substances could be the "candidates" for realization of all
three types of long-range order - ferromagnetic (due to RKKY
interaction), ferroelectric (due to carrier mediated dipole-dipole
interaction, see, e.g., Ref.\cite{vlad}) and ME order, also
carrier-mediated. This means that the interaction via charge
carriers can make the above substance a multiferroic. Note, that phases with ME
order belong to secondary ferroics with order parameter involving
both magnetization and polarization \cite{glin_mor08}, their domain
state can be switched by simultaneous application of magnetic and
electric fields \cite{wa}.

If local spins and electric dipoles (off-center ions) are randomly
distributed over the host lattice sites, the overall thermodynamic
behavior depends on the history of the sample i.e. heating and/or
cooling. These effects, characteristic for disordered systems
\cite{klem_obz}, have been observed in ferroics with a high ME
effect \cite{nat04}.

In the present paper we show that the microscopic nature of the
magnetoelectric (ME) effect might be the interaction between local
spins and dipoles via free charge carriers in a mixed valence oxide.
In other words, if the substance under consideration has free
carriers, they can mediate the ME effect. In the mean field
approximation, we calculated for the first time the magnitude of carrier mediated ME
interaction and compared it to both conventional RKKY interaction and
dipole - dipole one. We obtain that the ME interaction oscillates
with the ratio $k_F d$ ($k_F$ is a Fermi wave vector, $d$ is
off-center ion displacement) and find the optimal value of $k_F$
(and hence the free electrons concentration $n_e$) when the ME
interaction magnitude is the same as that of RKKY interaction. The
dipole - dipole interaction magnitude is always smaller then both
RKKY and ME. This permits to hope that in the substances, where the
RKKY interaction between local spins is not small, one can also
achieve substantial value of a ME interaction along with
multiferroic state.

The global symmetry of the linear ME effect implies the {\em
simultaneous } breaking of time-reversal and inversion (say
"space-reversal") symmetries of a substance. To break both above
symmetries simultaneously, one needs to have in a system both spin
moments ${\vec S}$ (axial vectors, changing their directions at time
reversal) and dipole moments ${\vec d}$ (in other words, two level
system (TLS), see, e.g. \cite{vlad}), changing their directions at
spatial coordinates direction reversal. For a microscopic
description this means that a system Hamiltonian should incorporate
the "usual" terms proportional to $\sum _{ij}J_{SS}({\vec
r}_{ij}){\vec S}_i{\vec S}_j$ and $\sum _{ij}J_{dd}({\vec
r}_{ij}){\vec d}_i{\vec d}_j$ (responsible for magnetic and electric
long-range order respectively) as well as coupling terms
proportional to $\sum _{ij}J_{Sd}({\vec r}_{ij}){\vec S}_i{\vec
d}_j$. In our model, the latter term is responsible for the ME
effect in a solid.

To account for above effect microscopically, we should find
explicitly the potentials of interaction $J_{AB}$ (A,B=S,d). The
ubiquitous method of calculation of the above interaction potentials
$J_{AB}$ is a perturbation theory. In this approach, we have two
"zeroth" Hamiltonians - free electrons, TLS'ses, or spins. The
interaction term might be dipole-dipole (dd), spin-spin (SS) or
dipole-spin (Sd) interaction. If this term is considered to be
small, we can apply perturbation theory and get the interaction
potential, as usually, in its second order. The details of
calculations of dd and SS interactions can be found in Refs.
\cite{vlad,glinkond} and \cite{abr,kit} respectively. Our aim is to
calculate Sd interaction. This will be done similarly to above
interactions, using the second order correction to the ground state
energy $E_{\rm AB}^{(2)}$ (A,B=S,d). This expression reads
\cite{abr}
\begin{equation}\label{abr1}
    E_{\rm AB}^{(2)}=\sum _{ik}J({\bf r}_{ik}){\bf S}_i{\bf S}_k,
\end{equation}
where for finite temperatures
\begin{eqnarray}\label{abr2}
&&J({\bf r})=\sum_{{\bf k}_1{\bf k}_2}\frac{|U({\bf k}_1-{\bf
k}_2)|^2n_{{\bf k}_1}(1-n_{{\bf k}_2})}{\xi _{{\bf k}_1}-\xi _{{\bf
k}_2}}\times \nonumber \\
&&\times \exp[i({\bf k}_1-{\bf k}_2){\bf r}],\ n_{\bf
k}=\frac{1}{\exp\left(\frac{\xi ({\bf k})-\mu}{T}\right)+1}.
\end{eqnarray}
Here $U({\bf k}_1-{\bf k}_2)$ is the corresponding interaction
amplitude, $\mu $ is a chemical potential. Actually, the specific
form of $J({\bf r})$ (i.e. dd, SS or Sd) depends on the form of
$U({\bf k}_1-{\bf k}_2)$. Here $\xi ({\bf k})$ is a dispersion law
for free electrons
\begin{equation}\label{dl}
\xi ({\bf k})=\frac{\hbar^2k^2}{2m^*},
\end{equation}
and $n_{{\bf k}_i}$ are Fermi-Dirac factors.

Although the finite temperature effects might be important, it can
be shown that all important physics comes from the zero temperature
case. This means that we may safely use the zero temperature
approximation, which is actually used, for example, in RKKY
interaction derivation \cite{kit}. At zero temperature the
occupation numbers are simply unit step functions $\theta (\xi ({\bf
k})-\epsilon _f)$, where $\mu (T=0)=\epsilon _f$ is Fermi energy.

The coupling function $U({\bf k}_1-{\bf k}_2)$ is expressed via the
expansion over spherical harmonics of the angle $\theta_{{\bf k}_1,
{\bf k}_2}$ between vectors ${\bf k}_1$ and ${\bf k}_2$, see Ref.
\cite{vlad} for details. It can be shown that for our purposes it is
sufficient to consider the function $U$ to be isotropic, i.e. to use
the zeroth term of this expansion
\begin{equation}\label{psev1}
    U({\bf k}_1-{\bf k}_2)\equiv a_0\ \phi({\bf k}_1-{\bf k}_2)
\end{equation}
and the function $\phi ({\bf k}_1-{\bf k}_2)$ actually determines
the kind of interaction (SS,dd or Sd).

The spin-spin term $J_{SS}$ is indeed a well-known RKKY interaction,
in this case
 $\phi _{\rm SS} ({\bf k}_1-{\bf k}_2) \equiv i$ and

\begin{equation}\label{rkk2}
J({\bf r})=\frac{mV^2a_0^2k_f^4}{2\pi ^3\hbar ^2}\ \frac{\sin
x-x\cos x}{x^4}, x\equiv 2k_fr,
\end{equation}
where $V$ is a crystal volume.

For calculation of dipole - dipole coupling the interaction function
should be chosen in the form \cite{vlad}
$\phi _{\rm dd}({\bf k}_1-{\bf k}_2)=i\sin\left[\frac 12 {{\bf
d}}({\bf k}_1-{\bf k}_2)\right]$.
Substitution of this function into Eq. \eqref{abr2} at zero
temperature after a straightforward but lengthy calculations yields
\begin{eqnarray}\label{jxv}
J(r)&=&-\frac{mV^2a_0^2 k_f^4}{8\pi ^3\hbar ^2}\left[2F(x)-F(\psi
_+)-F(\psi_-)\right],\nonumber \\
F(y)&=&\frac{\sin y-y\cos y}{y^4},\ x=2k_fr,\ \psi
_{\pm}=a_{\pm}x,\nonumber \\
a_\pm &\equiv& \sqrt{1+\lambda ^2\pm 2\lambda \cos \alpha},\ \lambda
=\frac{d}{r}, d\equiv|{\bf d}|,
\end{eqnarray}
$\alpha$ is the angle characterizing the direction of vector ${\bf
d}$ (for example, it can be thought as the angle between ${\bf d}$
and z axis. It is well-known (see above) that the off-center ions
can have several orientations in a host lattice. The different
orientations mean different values of angle $\alpha$. For example,
for a two position orientable dipole $\alpha=0,\pi$. This fact will
be used below.

The Sd interaction potential will also depend on the angle $\alpha$,
its function $|U({\bf k}_1-{\bf k}_2)|^2=a_0^2(\phi _{\rm SS}^*\phi
_{\rm dd}+c.c.)$ (c.c. stands for complex conjugated). This yields
\begin{eqnarray}\label{fn}
&&J_{\rm Sd}({\bf r})=-\frac{ma_0^2V^2k_f^4}{4\pi ^3\hbar
^2}[F(\eta _-)-F(\eta _+)],\nonumber \\
&&\eta _\pm=b_\pm x,\ b_\pm=\sqrt{1+\frac 14 \lambda ^2\pm \lambda
\cos\alpha}.
\end{eqnarray}
The expression (\ref{fn}) gives the desired potential of spin-dipole
interaction mediated  by charge carriers.

To estimate the magnitude of the ME interaction, we calculate Weiss
mean fields, related to SS, Sd and dd interactions. Having the
interaction $J_{\rm AB}({\bf r})$ ($A,B=S,d$), the Weiss field is
defined as
\begin{equation}\label{wf1}
    W_{\rm AB}=\frac{1}{V}\int _V J_{\rm AB}({\bf r})d^3r,
\end{equation}
where $V$ is the crystal volume. Since all $J_{\rm AB}({\bf r})$
except RKKY \eqref{rkk2} depend on ${\bf d}$ - the ion off-central
shift, the resulting integrals will be oscillating functions of this
parameter (to be more specific, the parameter is $k_Fd$). This means
that depending on $k_Fd$ value, the corresponding Weiss fields may
be positive or negative, defining the mutual direction of spins and
dipoles in ferroelectric and magnetoelectric phases.

We recollect here, that in a mean field approximation the
temperatures of the transition to ferromagnetic, ferroelectric and
magnetoelectric phases are related to corresponding Weiss fields in
an obvious fashion
    $T_{c\rm{AB}}=\sqrt{n_A n_B}|W_{\rm AB}|,$
where $n_{A,B}$ is a concentration (number of particles per unit
volume) of local spins or electric dipoles.

\begin{figure}
\hspace*{-9mm}
\includegraphics [width=0.54\textwidth]{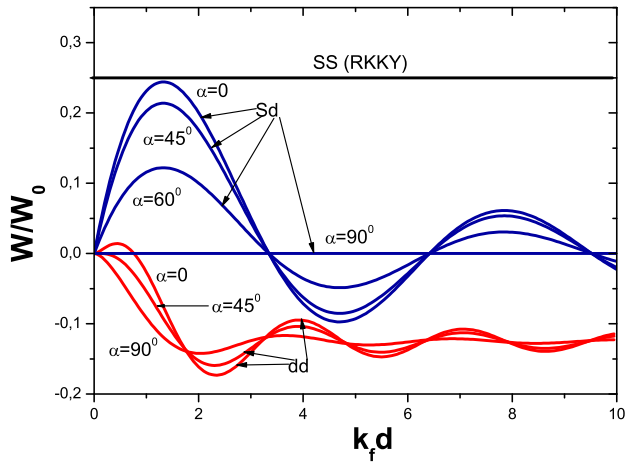}
\vspace*{-1.0cm} \caption{The magnitudes of SS, Sd and dd
interactions (Weiss fields) as functions of parameter $k_Fd$ at
different directions of the off-center shift, defined by angle
$\alpha $.}
\end{figure}

The expressions \eqref{jxv} and \eqref{fn} show that the
corresponding integrals cannot be done analytically. Substitution of
the RKKY potential \eqref{rkk2} into Eq. \eqref{wf1} yields

\begin{equation}\label{wf2}
    W_{\rm SS}=\frac 14 W_0,\ W_0=\frac{m^*Va_0^2}{\pi ^2 \hbar ^2}.
\end{equation}

The estimate for $W_0$ here gives $W_0\approx 500 K$ \cite{vlad}. In
that case $W_{\rm SS}\approx 125 K$. The results of numerical
calculation of $W_{\rm Sd}$ (ME interaction) and $W_{\rm dd}$
(dipole-dipole interaction) are reported in Fig.1 at different
angles $\alpha$. It is seen that while the magnitude of the magnetic
interaction $W_{\rm SS}$ is constant, the magnitudes of $dd$ and
$Sd$ interactions oscillate with the parameter $k_Fd$. Moreover
$|W_{\rm dd}|<$ $|W_{\rm Sd}|<$ $W_{\rm SS}$ and at $\alpha =0$ and
$k_Fd\approx 1.4$ the value of $|W_{\rm Sd}|$ is maximal, it is
almost equal to $W_{\rm SS}=0.25W_0$ and is much more then $|W_{\rm
dd}|$. Actually, the ME effect can be realized in broader interval
(see MD curve for $\alpha=0$ on Fig.1) $|W_{\rm Sd}|>0.5|W_{\rm
SS}|$ so that $0.5\leq k_Fd\leq 1.4$. This means that for a
substance, having oxygen vacancies (which is a source of free
carriers) and ions with off-center shift $d$ there is strong
possibility for substantial ME coupling via free carriers.

We estimate carriers concentration in the above $k_F$ interval with the
help of expression $n_e= \sqrt{n_0n_V}\exp(- E_f/T)$, valid for low
temperatures $T \leq 100$ K\cite{anselm}, where $n_0\approx
(10^{22}-10^{23})$cm$^{-3}$ is high temperature value of carrier
concentration, $n_V\approx 10^{17}$cm$^{-3}$ is oxygen vacancies
concentration, $E_f=\hbar^2k_f^2/(2m^*)$ is Fermi energy and $m^*$
is carrier effective mass. For $d = 0.5\cdot10^{-8}$ cm, $m^* = 10
m_e$ we obtain $n_e =(10^{16}-10^{17})$ cm$^{-3}$, which is a
typical value for semiconductors.

If the above substances have other (then oxygen vacancies) sorts of
defects, which can supply the free charge carriers, the above
interactions are stronger, promoting the multiferroic behavior at
low temperatures ($T \leq 100$ K). Note, that many single-phase
multiferroics exhibit multiferroicity only at low temperatures
\cite{nat04}. Our analysis shows (see Refs
\cite{stef97,semst02,semst03}) that the randomness in the above ion
positions as well as presence of other types of unavoidable
technological defects in a substance plays a dual role. On one hand,
it tries to destroy all three kinds of above long-range order, the
most sensitive ferroelectric one, which is due to dipole-dipole
interaction. On the other hand, this disorder might supply extra
free electrons or holes, which promote the considered effects.
Moreover, the long-range magnetic order (due to RKKY interaction) is
least sensitive to disorder and ME order is an intermediate case.
Thus we come to the conclusion that disorder in the above substances
would not adversely influence initial ME order.

It is also seen from the Fig.1 that since RKKY interaction is
dominating in the problem under consideration, the other important
factor needed for the ME effect to be realized is a sufficient
strength of spin-spin (RKKY) interaction via free carriers. If $T_c$
related to RKKY interaction is around 150 K, then (see Fig.1) the ME
phase transition temperature $T_{cME}$ will be of the same
magnitude. Moreover, the maximal value of $T_{cME}$ occurs for
$\alpha =0$, i.e. for a two site orientable dipole. For other angles
$\alpha$ and parameters $k_Fd$ the situation is also not that
critical, the only "forbidden" values $k_Fd$ are those where $W_{\rm
Sd}=0$, $\alpha=90^\circ$ (see Fig.1). This shows that there is
large space of parameter values where the ME effect mediated by free
carriers can be realized in the above substances.

To conclude, we have shown that it is quite probable to obtain a
sufficiently strong ME coupling mediated by free charge carriers in
mixed valency oxides. Namely, in the substances, where RKKY
interaction between local spins is not small, and the free carrier
concentration is related to the off-center shift, a substantial ME
coupling may be realized. Smaller (but still sensitive) values of
above coupling occur in the entire domain of $k_F d$ values except
those, where $W_{\rm Sd}=0$ (see Fig.1). Our preliminary analysis
shows, that the effects of disorder influence multiferroic behavior
in the above substances but do not suppress it. The results of
quantitative consideration of disorder effects leading, for example,
to the dependence of ME characteristics on the sample history
\cite{hist} will be published elsewhere.
\begin{acknowledgments}
This work was supported in part by MULTICERAL grant.
\end{acknowledgments}

\end{document}